\documentclass[aps,prd,twocolumn,showpacs,amsmath,amssymb,bm,superscriptaddress]{revtex4-1}

\usepackage{graphicx}  
\usepackage{dcolumn}   
\usepackage{hyperref}
\usepackage{bm}        
\usepackage{float}
\newcommand{\amuD}{\ensuremath{A^{\mu D^{0}}}}

\newcommand {\asld} {\ensuremath{a^d_{\mathrm{sl}}}}
\newcommand {\asls} {\ensuremath{a^s_{\mathrm{sl}}}}
\newcommand {\aslq} {\ensuremath{a^q_{\mathrm{sl}}}}

\newcommand {\ks} {\ensuremath{K^0_S}}

\newcommand {\Bd} {\ensuremath{B^0_d}}
\newcommand {\Bs} {\ensuremath{B^0_s}}
\newcommand {\Bq} {\ensuremath{B^0_q}}

\newcommand {\barBq} {\ensuremath{\bar{B}^0_q}}

\newcommand{\dzero}     {\ensuremath{ D^0} \,}
\newcommand{\adzero}     {\ensuremath{\overline{ D}^0 \,}}

\begin{document}

\widetext

\hspace{5.2in} \mbox{FERMILAB-PUB-16-300-E}

\title{
Measurement of  the direct CP violating charge asymmetry  in 
$\bm{B^\pm \rightarrow \mu^\pm \nu_\mu D^{0}}$   decays}

\date{2 August 2016, resubmitted 10 January 2017}

\begin{abstract}
	We present the first measurement of the CP violating charge
asymmetry in $B^\pm \to \mu^\pm \nu_\mu {D}^0$ decays using the full
Run II integrated luminosity of 10.4 fb$^{-1}$ in proton-antiproton
collisions collected with the D0 detector at the Fermilab Tevatron
Collider. We measure a  difference in the yield of $B^-$ and  $B^+$
mesons in these decays  by fitting   the reconstructed invariant mass
distributions. This results  in an  asymmetry of $\amuD =\left[ -0.14
\pm 0.20 \right]\%$, which is consistent with  standard model
predictions.
\end{abstract}

\affiliation{LAFEX, Centro Brasileiro de Pesquisas F\'{i}sicas, Rio de Janeiro, RJ 22290, Brazil}
\affiliation{Universidade do Estado do Rio de Janeiro, Rio de Janeiro, RJ 20550, Brazil}
\affiliation{Universidade Federal do ABC, Santo Andr\'e, SP 09210, Brazil}
\affiliation{University of Science and Technology of China, Hefei 230026, People's Republic of China}
\affiliation{Universidad de los Andes, Bogot\'a, 111711, Colombia}
\affiliation{Charles University, Faculty of Mathematics and Physics, Center for Particle Physics, 116 36 Prague 1, Czech Republic}
\affiliation{Czech Technical University in Prague, 116 36 Prague 6, Czech Republic}
\affiliation{Institute of Physics, Academy of Sciences of the Czech Republic, 182 21 Prague, Czech Republic}
\affiliation{Universidad San Francisco de Quito, Quito, Ecuador}
\affiliation{LPC, Universit\'e Blaise Pascal, CNRS/IN2P3, Clermont, F-63178 Aubi\`ere Cedex, France}
\affiliation{LPSC, Universit\'e Joseph Fourier Grenoble 1, CNRS/IN2P3, Institut National Polytechnique de Grenoble, F-38026 Grenoble Cedex, France}
\affiliation{CPPM, Aix-Marseille Universit\'e, CNRS/IN2P3, F-13288 Marseille Cedex 09, France}
\affiliation{LAL, Univ. Paris-Sud, CNRS/IN2P3, Universit\'e Paris-Saclay, F-91898 Orsay Cedex, France}
\affiliation{LPNHE, Universit\'es Paris VI and VII, CNRS/IN2P3, F-75005 Paris, France}
\affiliation{CEA Saclay, Irfu, SPP, F-91191 Gif-Sur-Yvette Cedex, France}
\affiliation{IPHC, Universit\'e de Strasbourg, CNRS/IN2P3, F-67037 Strasbourg, France}
\affiliation{IPNL, Universit\'e Lyon 1, CNRS/IN2P3, F-69622 Villeurbanne Cedex, France and Universit\'e de Lyon, F-69361 Lyon CEDEX 07, France}
\affiliation{III. Physikalisches Institut A, RWTH Aachen University, 52056 Aachen, Germany}
\affiliation{Physikalisches Institut, Universit\"at Freiburg, 79085 Freiburg, Germany}
\affiliation{II. Physikalisches Institut, Georg-August-Universit\"at G\"ottingen, 37073 G\"ottingen, Germany}
\affiliation{Institut f\"ur Physik, Universit\"at Mainz, 55099 Mainz, Germany}
\affiliation{Ludwig-Maximilians-Universit\"at M\"unchen, 80539 M\"unchen, Germany}
\affiliation{Panjab University, Chandigarh 160014, India}
\affiliation{Delhi University, Delhi-110 007, India}
\affiliation{Tata Institute of Fundamental Research, Mumbai-400 005, India}
\affiliation{University College Dublin, Dublin 4, Ireland}
\affiliation{Korea Detector Laboratory, Korea University, Seoul, 02841, Korea}
\affiliation{CINVESTAV, Mexico City 07360, Mexico}
\affiliation{Nikhef, Science Park, 1098 XG Amsterdam, the Netherlands}
\affiliation{Radboud University Nijmegen, 6525 AJ Nijmegen, the Netherlands}
\affiliation{Joint Institute for Nuclear Research, Dubna 141980, Russia}
\affiliation{Institute for Theoretical and Experimental Physics, Moscow 117259, Russia}
\affiliation{Moscow State University, Moscow 119991, Russia}
\affiliation{Institute for High Energy Physics, Protvino, Moscow region 142281, Russia}
\affiliation{Petersburg Nuclear Physics Institute, St. Petersburg 188300, Russia}
\affiliation{Instituci\'{o} Catalana de Recerca i Estudis Avan\c{c}ats (ICREA) and Institut de F\'{i}sica d'Altes Energies (IFAE), 08193 Bellaterra (Barcelona), Spain}
\affiliation{Uppsala University, 751 05 Uppsala, Sweden}
\affiliation{Taras Shevchenko National University of Kyiv, Kiev, 01601, Ukaine}
\affiliation{Lancaster University, Lancaster LA1 4YB, United Kingdom}
\affiliation{Imperial College London, London SW7 2AZ, United Kingdom}
\affiliation{The University of Manchester, Manchester M13 9PL, United Kingdom}
\affiliation{University of Arizona, Tucson, Arizona 85721, USA}
\affiliation{University of California Riverside, Riverside, California 92521, USA}
\affiliation{Florida State University, Tallahassee, Florida 32306, USA}
\affiliation{Fermi National Accelerator Laboratory, Batavia, Illinois 60510, USA}
\affiliation{University of Illinois at Chicago, Chicago, Illinois 60607, USA}
\affiliation{Northern Illinois University, DeKalb, Illinois 60115, USA}
\affiliation{Northwestern University, Evanston, Illinois 60208, USA}
\affiliation{Indiana University, Bloomington, Indiana 47405, USA}
\affiliation{Purdue University Calumet, Hammond, Indiana 46323, USA}
\affiliation{University of Notre Dame, Notre Dame, Indiana 46556, USA}
\affiliation{Iowa State University, Ames, Iowa 50011, USA}
\affiliation{University of Kansas, Lawrence, Kansas 66045, USA}
\affiliation{Louisiana Tech University, Ruston, Louisiana 71272, USA}
\affiliation{Northeastern University, Boston, Massachusetts 02115, USA}
\affiliation{University of Michigan, Ann Arbor, Michigan 48109, USA}
\affiliation{Michigan State University, East Lansing, Michigan 48824, USA}
\affiliation{University of Mississippi, University, Mississippi 38677, USA}
\affiliation{University of Nebraska, Lincoln, Nebraska 68588, USA}
\affiliation{Rutgers University, Piscataway, New Jersey 08855, USA}
\affiliation{Princeton University, Princeton, New Jersey 08544, USA}
\affiliation{State University of New York, Buffalo, New York 14260, USA}
\affiliation{University of Rochester, Rochester, New York 14627, USA}
\affiliation{State University of New York, Stony Brook, New York 11794, USA}
\affiliation{Brookhaven National Laboratory, Upton, New York 11973, USA}
\affiliation{Langston University, Langston, Oklahoma 73050, USA}
\affiliation{University of Oklahoma, Norman, Oklahoma 73019, USA}
\affiliation{Oklahoma State University, Stillwater, Oklahoma 74078, USA}
\affiliation{Oregon State University, Corvallis, Oregon 97331, USA}
\affiliation{Brown University, Providence, Rhode Island 02912, USA}
\affiliation{University of Texas, Arlington, Texas 76019, USA}
\affiliation{Southern Methodist University, Dallas, Texas 75275, USA}
\affiliation{Rice University, Houston, Texas 77005, USA}
\affiliation{University of Virginia, Charlottesville, Virginia 22904, USA}
\affiliation{University of Washington, Seattle, Washington 98195, USA}
\author{V.M.~Abazov} \affiliation{Joint Institute for Nuclear Research, Dubna 141980, Russia}
\author{B.~Abbott} \affiliation{University of Oklahoma, Norman, Oklahoma 73019, USA}
\author{B.S.~Acharya} \affiliation{Tata Institute of Fundamental Research, Mumbai-400 005, India}
\author{M.~Adams} \affiliation{University of Illinois at Chicago, Chicago, Illinois 60607, USA}
\author{T.~Adams} \affiliation{Florida State University, Tallahassee, Florida 32306, USA}
\author{J.P.~Agnew} \affiliation{The University of Manchester, Manchester M13 9PL, United Kingdom}
\author{G.D.~Alexeev} \affiliation{Joint Institute for Nuclear Research, Dubna 141980, Russia}
\author{G.~Alkhazov} \affiliation{Petersburg Nuclear Physics Institute, St. Petersburg 188300, Russia}
\author{A.~Alton$^{a}$} \affiliation{University of Michigan, Ann Arbor, Michigan 48109, USA}
\author{A.~Askew} \affiliation{Florida State University, Tallahassee, Florida 32306, USA}
% ...
%
%
\collaboration{The D0 Collaboration\footnote{with visitors from
$^{a}$Augustana College, Sioux Falls, SD 57197, USA,
$^{b}$The University of Liverpool, Liverpool L69 3BX, UK,
$^{c}$Deutshes Elektronen-Synchrotron (DESY), Notkestrasse 85, Germany,
$^{d}$CONACyT, M-03940 Mexico City, Mexico,
$^{e}$SLAC, Menlo Park, CA 94025, USA,
$^{f}$University College London, London WC1E 6BT, UK,
$^{g}$Centro de Investigacion en Computacion - IPN, CP 07738 Mexico City, Mexico,
$^{h}$Universidade Estadual Paulista, S\~ao Paulo, SP 01140, Brazil,
$^{i}$Karlsruher Institut f\"ur Technologie (KIT) - Steinbuch Centre for Computing (SCC),
D-76128 Karlsruhe, Germany,
$^{j}$Office of Science, U.S. Department of Energy, Washington, D.C. 20585, USA,
$^{k}$American Association for the Advancement of Science, Washington, D.C. 20005, USA,
$^{l}$Kiev Institute for Nuclear Research (KINR), Kyiv 03680, Ukraine,
$^{m}$University of Maryland, College Park, MD 20742, USA,
$^{n}$European Orgnaization for Nuclear Research (CERN), CH-1211 Geneva, Switzerland
and
$^{o}$Purdue University, West Lafayette, IN 47907, USA.
$^{\ddag}$Deceased.
}} \noaffiliation
\vskip 0.25cm

\pacs{13.25.Hw, 11.30.Er,  14.40.Nd}
\maketitle

\section{Introduction}
Direct CP violation (CPV) in the semileptonic decay $B^+ \to
\mu^+ \nu_\mu \overline{D}^0$ does not occur  in  the standard model
(SM).  Any CPV in
this decay would indicate the existence of non-SM physics. The
anomalously large CP-violating effects in the like-sign dimuon
asymmetry measured by the D0 Collaboration~\cite{dimuon2013} could be
explained by the presence of direct CPV in semileptonic decays. This
article presents the first  measurement of the direct CP-violating
charge asymmetry.   We use  the full Run II integrated luminosity  of
10.4 fb$^{-1}$ of proton-antiproton collisions collected with the D0
detector at the Fermilab Tevatron Collider. Charge conjugate states are assumed in this paper.

The CPV charge asymmetry is defined as 
\begin{equation}
\amuD = \frac{\Gamma(B^-\to \mu^-\overline{\nu}_{\mu}D^0) - \Gamma(B^+\to \mu^+{\nu_\mu}\overline{D}^0)}
{\Gamma(B^-\to \mu^-\overline{\nu}_{\mu}D^0) + \Gamma(B^+\to \mu^+{\nu_\mu}\overline{D}^0)}.
\end{equation}
We assume that there is no production asymmetry between $B^+$ and
$B^-$ mesons in proton-antiproton collisions and we estimate that any
production asymmetry of $b$ baryons and other $B$ mesons that decay
to $\mu^+\overline{D}^0$ is small (see below for further discussion).
The measurement is performed using the raw asymmetry 
\begin{equation}
\label{eq:raw}
A_{\rm{raw}} = \frac{N_{\mu^-D^0} - N_{\mu^+\overline{D}^0}}{N_{\mu^-D^0} + N_{\mu^+\overline{D}^0}},
\end{equation}
where $N_{\mu^-D^0}$ ($N_{\mu^+\overline{D}^0}$) is the number of
reconstructed $B^- \to \mu^- \overline{\nu}_\mu {D}^0$ ($B^+ \to \mu^+
\nu_\mu \overline{D}^0$) decays. This includes all decay processes of
$B^+$ mesons that result in a $D^0$ meson and an appropriately charged
muon in the final state. Neglecting any terms that are second or higher
order in the asymmetry the charge asymmetry in $B^\pm$ decays is  given
by
\begin{equation}
A_{\rm{raw}}
 =  f_{B^{+}} \amuD + A_{\rm{det}} + A_{\rm{phys}},
\end{equation}
where $f_{B^+}$ is  fraction of the $\mu^+ \nu_\mu \overline{D}^0$
events produced by the decay of a $B^+$ meson, $A_{\rm{det}}$ is due to 
reconstruction asymmetries in the detector, and $A_{\rm{phys}}$ is the
charge asymmetry resulting from the decay of other particles in the
sample.

\section{Data Selection}

The D0 detector has a central tracking system consisting of a silicon
microstrip tracker (SMT) and the central fiber tracker (CFT), both
located within a 2~T superconducting solenoidal magnet~\cite{d0det,
layer0}. A muon system, covering $|\eta|<2$~\cite{eta}, consists of a
layer of tracking detectors and scintillation trigger counters in front
of 1.8~T toroidal iron magnets, followed by two similar layers after the
toroids~\cite{run2muon}.

The polarities of the toroidal and solenoidal magnetic fields are
reversed on average every two weeks so that the four solenoid-toroid
polarity combinations are exposed to approximately the same integrated
luminosity. This allows for a cancellation of first-order effects
related to instrumental charge and momentum reconstruction asymmetries.
To ensure a more complete cancellation of the uncertainties, the events
are weighted according to the number of 
$\mu^+ \nu_\mu \overline{D}^0$ decays
collected in each configuration of the magnet polarities (polarity
weighting). The weighting is based on the number of events containing
$D^0$ decay products that pass the selection criteria and the likelihood
selection (described below), and that are in the $K^+\pi^-$ invariant
mass range used for the fit.

The data are collected with a suite of single and dimuon triggers. $B^+$
mesons are selected using their semileptonic decays $B^+ \to \mu^+
\nu_\mu \overline{D}^0$ by applying criteria similar to those used in
Ref.~\cite{Bdmixing}. Muons are  required to have hits in more than one
muon chamber, an associated track in the central tracking system with
hits in both SMT and CFT, transverse momentum $p_T^{\mu} > 2$ GeV/$c$ as
measured in the central tracker, pseudorapidity $|\eta^\mu|<2$, and
total momentum $p^{\mu} > 3$ GeV/$c$. The muons that satisfy the selection 
criteria pass through 12.8 to 14.5 hadronic interaction lengths. The  background from hadrons 
faking muons is negligible. 

All charged particles in a given event are clustered into jets using the
DURHAM clustering algorithm \cite{durham} with the cut-off parameter set
to 15 GeV/$c$. Events with more than one identified muon in the same jet
or with a reconstructed $J/\psi \to \mu^+ \mu^-$ decays are rejected.

\adzero candidates are constructed from two tracks of opposite sign of 
curvature associated  with the same jet as the reconstructed muon. Both
tracks are required to have transverse momentum of $p_T > 0.7$ GeV/$c$.
They are required to form a common  vertex with a fit $\chi^2 < 16$ for which
the number of degrees of freedom (ndof) is 1. The distance $d_T^D$
between the $p\bar{p}$ collision and \adzero\ vertices in the transverse
plane is required to exceed 3 standard deviations, $d_T^D/\sigma(d_T^D)
> 3$. The tracks of the muon and \adzero\ candidate must form a common 
vertex with a fit $\chi^2 < 16$ (ndof = 1). The mass of the kaon is
assigned to the track having the same sign of curvature as the muon. The
remaining track is assigned the mass of the charged pion. The mass of
the $\mu^+ \adzero$ system is required to be $2.0 < M(\mu^+ \adzero) <
5.5$ GeV/$c^2$. The distance $d_T^B$ between the $p\bar{p}$ collision
and $B$ vertices in the transverse plane must be $>3\sigma(d_{T}^{B})$.

\begin{figure}[thb]
\begin{center}
\vspace*{5mm}

\includegraphics[width=\columnwidth]{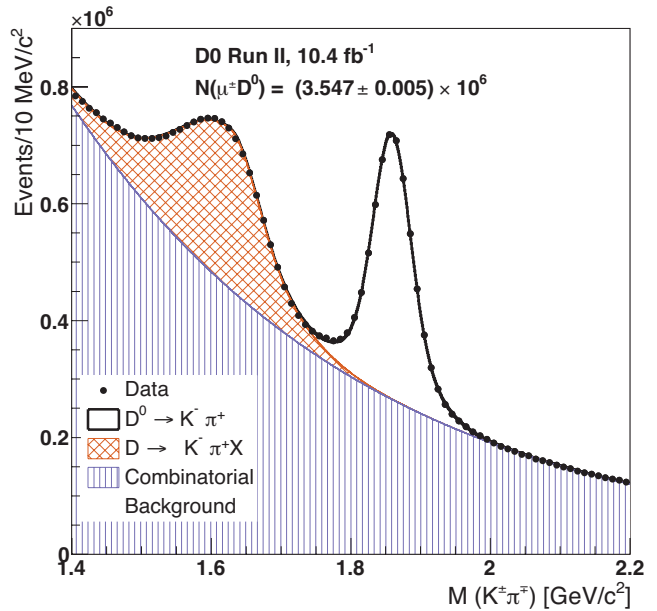} \\
\caption{\label{fig:fig1}
   The sum of the $K^+\pi^-$ and $K^-\pi^+$ invariant mass distributions
   for selected $\mu \dzero$ candidates. The curve shows the result of
   the fit described in the text.
}
\end{center}
\end{figure}

The $K^\pm \pi^\mp$ mass distribution for the selected  sample  and the
results of the fit to signal and background components are shown in Fig.
\ref{fig:fig1}. 

\vspace*{1mm}

\section{Raw Asymmetry}

We choose a fitting function to give  a good representation of the $K^+
\pi^-$ mass spectrum over the entire sample of $\mu^+ \nu_\mu\adzero$
polarity weighted events shown in Fig. \ref{fig:fig1}. The signal peak
corresponding to the decay $\overline{D}^0 \to K^+ \pi^-$ lies at
$M(K^\pm \pi^\mp) =  1.857$ GeV/$c^2$. The background in the mass region
above the signal  is adequately described by an exponential function in
the $K\pi$ mass $M$:
\begin{equation}
f_{1}^{\text {bkg}}(M)   =  \exp\left({a_0-a_1 M}\right),
\label{funcbkg_1}
\end{equation}
where $a_0$ and $a_1$ are fit parameters.

The signal is modelled by the sum of two Gaussians:
\begin{align}
\label{funcsig}
f^{\text {sig}} (M)  = & \frac{N^{\text {sig}}}{\sqrt{2 \pi}}  
 \left[ \frac{r_1}{\sigma_1} 
\exp\left({-\frac{(M-M_{D^0})^2}{2 \sigma_1^2}}\right) \right . \nonumber \\ 
  & \left. +     \frac{1-r_1}{\sigma_2} 
 \exp\left({-\frac{(M-M_{D^0})^2}{2 \sigma_2^2}}\right) \right],
\end{align}
where $N^{\text {sig}}$ is the number of signal events, $M_{D^0}$ is the
mean of the Gaussian functions, $\sigma_1$ and $\sigma_2$ are their
widths, and $r_1$ is the fractional contribution of the first Gaussian
function.

The peak in the background below the signal region is due to $D$ mesons
decaying to  $K^+\pi^- X$, where $X$ is not reconstructed ($X$ is
typically a $\pi^0$). It is modelled with a bifurcated Gaussian function:
\begin{widetext}
\begin{align}
f_{2}^{\text{bkg}}(M)   = & N_2 \left[ r_1 \exp \left({-\frac{(M-\mu_0)^2}{2 \sigma_R^{2}}} \right) + (1-r_1) \exp \left({-\frac{(M-\mu_0)^2}{2 (S\sigma_R)^{2}}} \right)\right]
\mbox{~~for~~} M-\mu_0 \geq 0,  \nonumber \\ 
  = & N_2 \left[ r_1 \exp \left({-\frac{(M-\mu_0)^2}{2 \sigma_L^{2}}} \right) + (1-r_1) \exp \left({-\frac{(M-\mu_0)^2}{2 (S\sigma_L)^{2}}} \right)\right]
\mbox{~~for~~} M-\mu_0 < 0. 
\label{funcbkg_2}
\end{align}
\end{widetext}
Here, $\mu_0$ is the mean of the Gaussian function,  $\sigma_L$ and
$\sigma_R$ are the two widths of the bifurcated Gaussian function, 
$r_1$ is the fractional contribution of the first Gaussian function
which is constrained to be the same as the fraction in the signal peak,
and $S = \sigma_2/\sigma_1$ from the fitted signal peak
(Eq.~\ref{funcsig}). These constraints are a result of the detector mass
resolution and are required for the fit to converge.

\begin{figure}[htbp]
\includegraphics[width=\columnwidth]{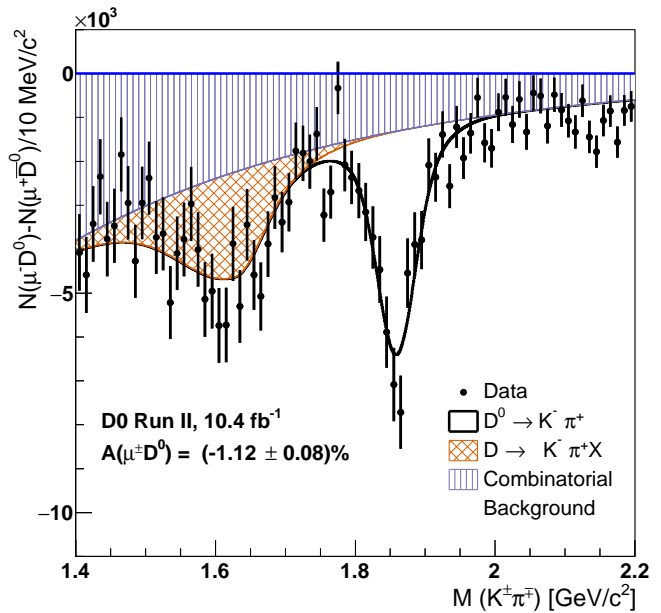}
\caption{\label{fig:fig4} 
	The results of a fit to the differences between the numbers of $\mu^-
	D^0$ and $\mu^+ \overline{D}^0$ events as function of the $K\pi$ mass.  }
\end{figure}

The fit yields  $(3.547 \pm 0.005) \times 10^6$ $\mu^+ \adzero$
candidates. The raw asymmetry (Eq.~\ref{eq:raw}) is extracted by fitting
the $K^+\pi^-$ mass spectrum using a $\chi^2$ minimization. The fit is
performed simultaneously, using the same model, on the sum
(Fig.~\ref{fig:fig1}) and the difference (Fig.~\ref{fig:fig4}) of the
$M(K^-\pi^+)$ distribution for the $\mu^- D^0$ candidates and the
$M(K^+\pi^-)$ distribution for the $\mu^+ \overline{D}^0$ candidates.
The functions used to model the two distributions are

\begin{align}
 \label{funcfit2}
f_{\rm{sum}}(M)  = &  f^{\text {sig}}(M) + f_{1}^{\text {bkg}}(M) + f_{2}^{\text{bkg}}(M), \\ 
f_{\rm{diff}}(M)  = &  A_{\rm{raw}} f^{\text {sig}}(M) + A_1 f_{1}^{\text {bkg}}(M) + A_2 f_{2}^{\text{bkg}}(M).
\end{align} 
Here, $A_1$ is the asymmetry of the
combinatoric background, and $A_2$ is the asymmetry of the $\overline{D}$ mesons that decay to $K^+\pi^- X$,
where $X$ is not reconstructed. The fitted asymmetry parameters are
$A_{\rm{raw}} = (-1.12 \pm 0.08)\%$, $A_1
= (-0.50 \pm 0.03)\%$ and $A_2 = (-0.87 \pm 0.12)\%$ (Fig.~\ref{fig:fig4}) where the uncertainties are
statistical.

Systematic uncertainties on the fitting method are evaluated by varying
the fitting procedure. The mass range of the fit is shifted from $1.40 <
 M < 2.20$~GeV$/c^2$ to $1.43 <  M < 2.17$~GeV$/c^2$ in steps of
10~MeV$/c^2$. The function modelling the high mass background is changed
to  a $2^{\rm nd}$ order polynomial function. The width of the mass bins
is varied between 5 and 20~MeV$/c^2$. The uncertainty for each of these
modifications to the fitting procedure is assigned to be half of the
maximal variation. The resulting  systematic uncertainty is $0.0075\%$
on the raw asymmetry $A_{\rm{raw}}$, yielding
\begin{equation}
A_{\rm{raw}} = \left[ -1.12 \pm 0.08 \thinspace (\text{stat}) \pm 0.008 \thinspace (\text{syst}) \right]\% .
\end{equation}

\section{Reconstruction Asymmetries} 
The residual reconstruction asymmetries  are described in
Ref.~\cite{d0adsl}. The residual detector tracking asymmetry has been
studied in Ref.~\cite{dimuon2010}  by using $\ks \rightarrow \pi^+\pi^-$
and $K^{\ast\pm} \rightarrow \ks \pi^\pm$ decays. These analyses show  
no significant residual track reconstruction asymmetries in the D0
detector. For this reason no  correction for tracking asymmetries is
applied ($A_{\text{track}} \equiv 0$). The reconstruction asymmetry of
charged pions has been studied using Monte Carlo (MC) simulations of the
detector \cite{dimuon2010}. The asymmetry is found to be less than
$0.05\%$ which is assigned as a systematic uncertainty and no correction
is made. The muon and the pion have opposite charge, so any remaining 
track asymmetries  will cancel   to first order.

The  residual reconstruction asymmetry of  muons  is measured using
$J/\psi \rightarrow \mu^+\mu^-$ decays as described in
\cite{dimuon2010,dimuon2}. This asymmetry is determined as a function of
$p_T(\mu)$ and $|\eta(\mu)|$, and the final correction is obtained by a
weighted average over the normalized ($p_T(\mu)$, $|\eta(\mu)|$) yields,
as determined from fits to the $M(K\pi)$ invariant mass distribution.
The resulting correction is
\begin{eqnarray} 
A_\mu = & \left[\epsilon(\mu^+) -\epsilon(\mu^-)\right]/\left[\epsilon(\mu^+) +\epsilon(\mu^-)\right] \nonumber \\
	= & \left[ 0.10 \pm 0.06 \thinspace (\mbox{syst})\right]\% ,
\end{eqnarray}
where $\epsilon(\mu^\pm)$ are the reconstruction efficiencies of
positively and negatively charged muons. This correction also includes
the systematic uncertainty due to track reconstruction.

The correction for a difference in behavior between positively and
negatively charged kaons is calculated using the measured kaon
reconstruction asymmetry presented in Ref.~\cite{d0adsl}. Negative kaons
can interact with matter to produce hyperons, while there is no
equivalent interaction for positive kaons. As a result, the mean path
length for positive kaons is larger, the reconstruction efficiency is
higher, and the kaon asymmetry $A_K$ is positive.

The kaon asymmetry is measured using a dedicated sample of $K^{\ast 0}
(\bar{K}^{\ast 0}) \to K^+\pi^- (K^- \pi^+)$ decays, based on the
technique described in Ref.~\cite{dimuon2010}. The $K^+\pi^-$ and
$K^-\pi^+$ signal yields are extracted by fitting the charge-specific
$M(K^{\pm}\pi^{\mp})$ distributions, and the asymmetry is determined by
dividing the difference by the sum of the distributions. The track
selection criteria in Ref.~\cite{d0adsl} are the same as those required
in the signal selection  in this analysis.

A strong dependence of the kaon asymmetry on kaon momentum  $p(K)$ and
the absolute value of the pseudorapidity $\eta(K)$ is found. Hence, the
final kaon asymmetry correction is determined from the polarity-weighted
average of $A_K[p(K),|\eta(K)|]$ over the $p(K)$ and $|\eta(K)|$
distributions in the signal events. A relative systematic uncertainty of
$5\%$ is assigned to each bin to account for possible variations in the
yield when different models are used to fit the signal and backgrounds
in the $K^{*0}$ mass distribution. Based on studies over a range of fit
variations the relative systematic uncertainty on the
$\mu^+\overline{D}^0$ yields per $p(K)$ and $|\eta(K)|$ bin is 1\%. The
resulting kaon asymmetry is found to be:
\begin{eqnarray}
A_K = & \left[\epsilon(K^+) -\epsilon(K^-)\right]/\left[\epsilon(K^+) +\epsilon(K^-)\right] \nonumber \\
 =  &\left[ 0.92 \pm 0.05 \thinspace (\mbox{syst})  \right] \%,
\end{eqnarray}
where $\epsilon(K^\pm)$ is the reconstruction efficiency of positively
and negatively charged kaons.

Combining the detector effects gives 
\begin{eqnarray}
A_{\rm{det}} = &  - A_{\mu} - A_{K} + A_{\text{track}} \nonumber \\
			 = & 
\left[ -1.02 \pm 0.08 \thinspace (\mbox{syst})  \right] \%.
\end{eqnarray}

\section{Signal Composition} 

The fraction of  $\mu^+ \nu_\mu \overline{D}^0$ events produced by the
decay of a $B^+$ meson, $f_{B^{+}}$, is extracted from MC  simulations.
The $\mu^+ \nu_\mu \overline{D}^0$ signal events can also be produced
via the decay  of \Bd\, mesons, \Bs\ mesons, $b$ baryon decays, and from
prompt $\overline{D}^0$ production. We generate a MC sample  using the
{\sc pythia}  event generator~\cite{pythia} modified to use {\sc
evtgen}~\cite{evtgen} for the decay of hadrons containing $b$ or $c$
quarks. The {\sc pythia}  inclusive QCD production model is  used.
Events recorded in random beam crossings are overlaid on the simulated
events to quantify the effect of additional collisions in the same or
nearby bunch crossings. Events are selected that contain at least one
muon and a  $\overline{D}^0 \to K^+\pi^-$ or ${D}^0 \to K^-\pi^+$ decay.
The generated events are processed by the full simulation chain, and
then by the same reconstruction and selection algorithms as used to
select events from real data. 

The mean proper decay lengths of  $b$ hadrons are fixed in the
simulation to values close to, but not exactly equal to, the current
world-average values~\cite{hfag}. To correct for these differences, an
event weighting is applied to all non-prompt events in the  simulation,
based on the generated lifetime of the $B$ candidate, to give the
world-average $B$ meson lifetimes. To estimate the effects of the
trigger selection and the reconstruction on the data, we weight each
event based on the transverse momentum of  the reconstructed muon. 
The combined  weighting applied to each MC event $i$ is given by $w_i$.
Combining all of these corrections,  we find  $f_{B^{+}} = 0.56$.

The remainder of the events in the signal are semileptonic decays 
of neutral $B$ mesons 
($B^0_{d} \rightarrow \mu^\pm D^{0}X$ and $B^0_{s} \rightarrow \mu^\pm D^{0}X$), 
 the combination of a muon and a $D^0$ from different sources including prompt 
 production  (combinatoric), and hadronic decays of $b$ hadrons where one of 
 the resulting hadrons decays semileptonically $h \to \mu\nu X$ ($B^\pm \rightarrow D^{0}h$, 
 $B^0_{d} \rightarrow  D^{0}h$, $B^0_{s} \rightarrow  D^{0}h$, and all other $b$-hadrons $\rightarrow D^0 h$). The sample composition is given 
 in Table~\ref{tab:signalcomp}. 

\begin{table}[h]
\caption{\label{tab:signalcomp}
Composition and mixing probability of the signal peak determined from MC simulation (the uncertainties are statistical).  The total  systematic uncertainty on  $f_{B^{+}}$  and the other signal fractions is
$0.01$. }
\begin{ruledtabular}
\newcolumntype{A}{D{A}{\pm}{-1}}
\begin{tabular}{lcr}
Decay Type & $P(\Bq \to \barBq)$  & \multicolumn{1}{c}{Fraction}\\
\hline
$B^\pm \rightarrow \mu^\pm D^{0}X$ 		 & n/a		& $(56.0 \pm 0.2)\%$ \\
$B^0_{d} \rightarrow \mu^\pm D^{0}X^\mp$ & 0.211	& $(35.2 \pm 0.2)\%$ \\
$B^0_{s} \rightarrow \mu^\pm D^{0}X^\mp$ & 0.5 		& $(1.8 \pm 0.1)\%$ \\
\hline 
Combinatoric & n/a & $(0.3 \pm 0.1)\%$ \\
$B^\pm \rightarrow D^{0}h$ & n/a & $(0.9 \pm 0.1)\%$ \\
$B^0_{d} \rightarrow  D^{0}h$ & 0.197	&	$(5.1 \pm 0.1)\%$ \\
$B^0_{s} \rightarrow  D^{0}h$ & 1.0 &  $(0 \pm 0.1)\%$ \\
Other $b$-hadrons  $\rightarrow  D^{0}h$ & n/a & $(0.7 \pm 0.1)\%$ \\ 
\end{tabular}
\end{ruledtabular}
\end{table}

To determine the systematic uncertainty on $f_{B^{+}}$, the exclusive branching
ratios and production fractions of $B$ mesons are varied by their
uncertainties, the $B$ meson lifetimes are varied within their
uncertainties, and  a coarser $p_T$ binning is used in the MC event weighting.
These variations are combined using a toy MC to determine the size of the systematic uncertainty for the simulation inputs. 
The total resulting systematic uncertainty on  $f_{B^{+}}$  and the other signal fractions is
$0.01$.

CP violation in the mixing of neutral $B$-mesons is a significant background in this analysis. 
These backgrounds depend on the fraction of neutral $B$-mesons 
that have oscillated into their antiparticle prior to decay, $\Bq \to \barBq$
or $\barBq \to \Bq$ where $q = d,s$. This fraction is given by 
\begin{equation}
P(\Bq \to \barBq) = \frac{1}{2W} \sum_i w_i \left[ 1 - \frac{\cos (\Delta m_q t)}{\cosh (0.5\Delta \Gamma_q t)}  \right] 
\end{equation}
where $\Delta m_q$, and $\Delta \Gamma_q$  are the  mass and decay rate differences of the mass eigenstates~[11], 
$w_i$ is the MC event weight and $W$ is the sum of the MC event weights.
The fractions for 
the different \Bd\ and \Bs\ samples are given in Table~\ref{tab:signalcomp}.
In the case of the
\Bs\ meson, the time-integrated oscillation probability is essentially 50\%  and is insensitive to the  
exact value of $\Delta m_s$.  The uncertainties on the mixing fraction are negligible when compared to 
the uncertainties on the sample composition.

\section{Physics Asymmetries}

The most significant potential contribution to   $A_{\rm{phys}}$  is
semileptonic charge asymmetries from the mixing of neutral $B$ mesons and is given by 
\begin{equation}
A_{\Bq} = \aslq P(\Bq \to \barBq) f_{\Bq}
\end{equation}
where $f_{\Bq}$ is the neutral meson signal fraction.
The world average \cite{hfag} semileptonic charge asymmetry from \Bd\
mixing is $\asld = (-0.15 \pm 0.17)\%$ which would lead to a
contribution to $A_{\rm{phys}}$ of $-0.011\%$.   The world average
\cite{hfag} semileptonic charge asymmetry for \Bs\ mixing is $\asls =
(-0.75 \pm 0.41)\%$ which would lead to a contribution to
$A_{\rm{phys}}$ of $-0.007\%$. Combining these asymmetries we obtain 
\begin{equation}
 A_{\rm{phys}} =   \left[ -0.02 \pm 0.02 \right]\%,
 \label{eq:aphys}
\end{equation}
where the systematic uncertainty is the combination of
uncertainties in the world averages combined with the uncertainties
on the sample composition. All other potential asymmetry
contributions are assumed to be negligible.

\section{Results} 

Combining the measured raw asymmetry,  and the detector and physics
corrections (Eqns 9, 12 and 15) and the estimated $B^+$ fraction 
$f_{B^{+}}$, we find
\begin{eqnarray}
  \amuD = & \left[ -0.14 \pm 0.14 \thinspace (\text{stat}) \pm 0.14 \thinspace (\text{syst}) \right]\%.  
\end{eqnarray}

We can estimate the size of a $B^\pm \rightarrow \mu^\pm \nu_\mu D^0$
asymmetry that would be needed to explain the observed like-sign dimuon
asymmetry~\cite{dimuon2013}. The like-sign dimuon asymmetry could be
explained by a semileptonic charge asymmetry in neutral $B$ mesons of
$A^B_{\mathrm{sl}} \sim 0.5\%$. A MC simulation of same sign dimuon
events where one muon originates from a neutral $B$-meson shows that
$62\%$ of these events also contain a muon from a semileptonic $B^+$
decay. Hence, $0.5\%/0.6 = 0.8\%$ would be  required to explain the
like-sign dimuon asymmetry. Thus our measurement implies that direct CPV
in $B^+$ decays is unlikely to  contribute a significant fraction of the
observed dimuon charge asymmetry, and that other explanations need to be
sought.

In summary, we have made  the first measurement of the direct
CP-violating charge asymmetry in $B^+$  mesons decaying semileptonically
to $\mu^+ \nu_\mu \overline{D}^0$. We find $\amuD =   \left[ -0.14 \pm
0.14 \thinspace (\text{stat}) \pm 0.14 \thinspace (\text{syst})
\right]\%$ where the total  uncertainty is 0.20\%. This result is in
agreement with the SM expectation of no CPV in this decay.

We thank the staffs at Fermilab and collaborating institutions,
and acknowledge support from the
Department of Energy and National Science Foundation (United States of America);
Alternative Energies and Atomic Energy Commission and
National Center for Scientific Research/National Institute of Nuclear and Particle Physics  (France);
Ministry of Education and Science of the Russian Federation, 
National Research Center ``Kurchatov Institute" of the Russian Federation, and 
Russian Foundation for Basic Research  (Russia);
National Council for the Development of Science and Technology and
Carlos Chagas Filho Foundation for the Support of Research in the State of Rio de Janeiro (Brazil);
Department of Atomic Energy and Department of Science and Technology (India);
Administrative Department of Science, Technology and Innovation (Colombia);
National Council of Science and Technology (Mexico);
National Research Foundation of Korea (Korea);
Foundation for Fundamental Research on Matter (The Netherlands);
Science and Technology Facilities Council and The Royal Society (United Kingdom);
Ministry of Education, Youth and Sports (Czech Republic);
Bundesministerium f\"{u}r Bildung und Forschung (Federal Ministry of Education and Research) and 
Deutsche Forschungsgemeinschaft (German Research Foundation) (Germany);
Science Foundation Ireland (Ireland);
Swedish Research Council (Sweden);
China Academy of Sciences and National Natural Science Foundation of China (China);
and
Ministry of Education and Science of Ukraine (Ukraine).


\begin{thebibliography}{99}
\bibitem{eta}
$\eta=-\ln[\tan(\theta/2)]\ $ is the pseudorapidity and $\theta$ is
the polar angle between the track momentum and the proton beam direction. $\phi$ is the azimuthal angle of the track.
\end{thebibliography}
\end{document}